\documentclass[aps,showpacs,twocolumn,amsmath,amssymb,superscriptaddress]{revtex4}
\usepackage{graphicx}
\usepackage{epstopdf}
\usepackage{bm}
\usepackage{subfigure}
\usepackage{sidecap}
\usepackage{times}
\usepackage{graphicx}
\usepackage{epstopdf}
\usepackage{amssymb}
\usepackage{amsmath}
\usepackage{dsfont,amsthm,amsbsy}
\usepackage{amssymb}
\usepackage{amsmath}
\usepackage{bbm}
\usepackage{graphicx}
\usepackage{epstopdf}
\usepackage{subfigure}
\usepackage{natbib}
\usepackage{epsfig}
\usepackage{amsfonts}
\usepackage{mathrsfs}
\usepackage{sidecap}
\usepackage{lipsum}
\usepackage{xcolor}
\usepackage[toc,page,title,titletoc,header]{appendix}
\usepackage[colorlinks,linkcolor=blue,citecolor=blue,anchorcolor=blue, urlcolor=blue]{hyperref}
\usepackage{hyperref}
\usepackage{resizegather}
\usepackage{float}
\usepackage{mathbbol}
\usepackage[normalem]{ulem}
\usepackage{cancel}
\usepackage{upgreek}
\usepackage{graphics,dcolumn,bm}
\usepackage{rotate,color}
\usepackage{times}
\usepackage{eqnarray}
\newcommand\redsout{\bgroup\markoverwith{\textcolor{red}{\rule[0.5ex]{2pt}{0.4pt}}}\ULon}

\newcommand{\bl}{\begin{aligned}}
\newcommand{\el}{\end{aligned}}
\def\be{\begin{equation}}
\def\ee{\end{equation}}

\def\bi{\begin{itemize}}
\def\ei{\end{itemize}}
\def\bn{\begin{enumerate}}
\def\en{\end{enumerate}}
\def\bea{\begin{eqnarray}}
\def\eea{\end{eqnarray}}
\def\no{\nonumber}
\def\ba{\begin{array}}
\def\ea{\end{array}}
\def\bd{\begin{displaymath}}
\def\ed{\end{displaymath}}

\def\tr{{\rm tr}}

\graphicspath{{Figures/}}
\begin{document}

\title{Engineering Floquet Dynamical Quantum Phase Transition}

\author{J. Naji}
\email[]{j.naji@ilam.ac.ir}
\affiliation{Department of Physics, Faculty of Science, Ilam University, Ilam, Iran}

\author{R. Jafari}
\email[]{jafari@iasbs.ac.ir, rohollah.jafari@gmail.com}
\affiliation{Department of Physics, Institute for Advanced Studies in Basic Sciences (IASBS), Zanjan 45137-66731, Iran}
\affiliation{School of physics, Institute for Research in Fundamental Sciences (IPM), P.O. Box 19395-5531, Tehran, Iran}
\affiliation{Department of Physics, University of Gothenburg, SE 412 96 Gothenburg, Sweden}
\affiliation{Beijing Computational Science Research Center, Beijing 100094, China}

\author{Longwen Zhou}
\email[]{zhoulw13@u.nus.edu}
\affiliation{College of Physics and Optoelectronic Engineering, Ocean University of China, Qingdao, China 266100}

\author{A. Langari}
\email[]{langari@sharif.edu}
\affiliation{Department of Physics, Sharif University of Technology, P.O.Box 11155-9161, Tehran, Iran}

\begin{abstract}
Floquet dynamical quantum phase transitions (FDQPTs) are signified by recurrent nonanalytic behaviors of observables in time. In this work, we introduce a quench-free and generic approach to engineer and control FDQPTs for both pure and mixed Floquet states. By applying time-periodic modulations with two
driving frequencies to a general class of spin chain model, we find multiple FDQPTs within each driving period.
The model is investigated with equal, commensurate and incommensurate driving frequencies.
The nonanalytic cusps of return probability form sublattice structures in time domain. Notably, the number and time-locations of these cusps can be flexibly controlled by tuning the Hamiltonian parameter and the frequencies of the drive. We further employ the dynamical topological order parameter (DTOP), which shows a quantized jump whenever a DQPT happens, to identify the topological feature of FDQPTs. Our findings reveal the advantage of engineering nonequilibrium phase transitions with multi-frequency driving fields.
\end{abstract}

\maketitle

\section{Introduction}
In recent decades, dynamical phase transitions -- phase transitions away from equilibrium -- have gained a lot of attention across many areas in the physics community, ranging from the abrupt changes in the relaxation dynamics of strongly correlated quantum many-particle systems \cite{Eckstein} to the domain formation in the early universe~\cite{Kibble}. The renaissance of the topic was commenced by the experimental advances achieved with ultracold atoms in optical lattices \cite{jotzu2014experimental,daley2012measuring,schreiber2015observation,choi2016exploring,flaschner2018observation,BlochRevModPhys2008}, making it possible to prepare and control nonequilibrium quantum states. Thereafter, trapped ions \cite{jurcevic2017direct,martinez2016real,neyenhuis2017observation,smith2016many}, nitrogen-vacancy center in diamonds \cite{yang2019floquet}, superconducting qubits \cite{guo2019observation} and photonic quantum walks \cite{wang2019simulating,xu2020measuring} were developed to provide a framework for experimentally studying a wide variety of dynamical phase transitions in nonequilibrium systems. These experiments have also provoked huge progress in theoretical physics.

Moreover, there has been growing interest in dynamical quantum phase transitions (DQPTs), which are
characterized theoretically by the nonanalyticity of physical observables in time domain. The notion of DQPTs was proposed as a counterpart of thermal phase transitions in equilibrium~\cite{Heyl2013,Heyl2018}. As the equilibrium phase transition is signalled by non-analyticities in the thermal free energy, the DQPT is revealed through the nonanalytical behavior of dynamical free energy, where the real time plays the role of a control parameter \cite{Heyl2013,Heyl2018,Jafari2019quench,andraschko2014dynamical,sedlmayr2018fate,Karrasch2013,vajna2014disentangling,
Jafari2019dynamical,Jafari2020,Mishra2018,Bhattacharya,Heyl2017,budich2016dynamical,Dutta2017,bhattacharya2017emergent,zhou2018dynamical,Sedlmayr2020,Longwen2021,
Abdi2019,Sadrzadeh2021,Yu2021,Modak2021,Lang2018,Peotta2021,Halimeh2017}. DQPTs, which were observed experimentally
\cite{flaschner2018observation,jurcevic2017direct,martinez2016real,guo2019observation,wang2019simulating,Nie2020,Tian2020}, display phase transitions between dynamically emerging quantum phases. They take place during the nonequilibrium coherent quantum time evolution under sudden/ramped quenches \cite{Heyl2013,Heyl2018,Jafari2019quench,andraschko2014dynamical,sedlmayr2018fate,Karrasch2013,vajna2014disentangling,
Jafari2019dynamical,Jafari2020,Mishra2018,Bhattacharya,Heyl2017,budich2016dynamical,Dutta2017,bhattacharya2017emergent,zhou2018dynamical,Sedlmayr2020,Longwen2021,
Abdi2019,Sadrzadeh2021,Yu2021,Modak2021,Divakaran2016,Sharma2016,Yu2022,Okugawa2022,Sedlmayr2022,Gonzalez2022,Hoyos2022,Brange2022,Hou2022,
Jensen2022,Rossi2022,Corps2022,Mondal2022,Stumper2022}
or time-periodic modulations of the Hamiltonian \cite{Zamani2020,Qianqian2021,Kosior2018a,Kosior2018b,yang2019floquet,Jafari2021,Zhou2019,Zhou2018,Jafari2022a,Naji2022a,Cai}. In addition, analogous to order parameters at equilibrium quantum phase transitions, dynamical topological order parameters (DTOPs) were proposed to capture the topological nature of DQPTs \cite{budich2016dynamical,Bhattacharya}. The DTOP is quantized and its unit magnitude jump at the critical time of DQPT reveals its topological feature \cite{budich2016dynamical,Bhattacharya,Zamani2020,Qianqian2021,yang2019floquet,Jafari2021,Zhou2019,Zhou2018}.

Since both the ground and excited states participate in the dynamics and the system keep exchanging energy with the driving field,
quantum many-body systems driven out of equilibrium via a periodic protocol yield exotic phenomena that are absent in those driven by a sudden or ramped quench. These include the generation of drive-induced topological states of matter \cite{Kitagawa2010,Kitagawa2011,Kundu2014,Mukherjee2018}, realization of Floquet time crystals \cite{Khemani2016,Else2016,zhang2017},
and phenomena such as dynamical localization \cite{Nag2014,Nag2015,Ghosh2020}, dynamical freezing \cite{Iubini2019,Divakaran2014}, and driving-induced tuning of ergodicity \cite{Mukherjee2020a,Mukherjee2020b}.
Consequently, studies of DQPTs in periodically driven systems -- known as Floquet DQPTs -- attracted a lot of attention.
It has been established that FDQPTs possess a class of DQPTs by displaying time-periodic non-analyticity and non-decaying return probabilities, which should make them easier to trace in the laboratory \cite{yang2019floquet,Zamani2020,Qianqian2021,Kosior2018a,Kosior2018b,Jafari2021,Zhou2019,Zhou2018}. Meanwhile, the conventional DQPTs following a single quench are usually observable only in transient time scales owing to the decaying return probabilities.

Therefore, realizing and controlling (effectively) closed non-equilibrium quantum many-body systems, specifically time-periodic driven systems, is of practical relevance as they might pave the way to the development of quantum technologies.
All the studies on controlling DQPTs and FDQPTs till now focus on sudden quench protocols, where the parameters of the given Hamiltonian are abruptly changed from one equilibrium phase to another \cite{Qianqian2021,Kosior2018a,Kosior2018b,Kennes2018}. One of the feature of equilibrium quantum phase transitions is the disability to adiabatically link the ground states between two distinct phases \cite{sachdevbook}. A nonanalyticity in the ground state energy is thus consistently encountered when crossing the critical point, irrespective of the path chosen to acquire this crossing. Therefore, controlling and engineering the time-periodic driven closed quantum systems in the context of Floquet theory without resorting to any quenches across the critical point is one of the most attractive topics in advancing nonequilibrium physics. Motivated by these considerations, we study the FDQPTs in a general class of periodically modulated model with two driving frequencies, in which one frequency guides the periodic evolution of the system, while the other frequency controls the FDQPTs therein.

In this paper, we elaborate on how the FDQPTs can be controlled simply by the Hamiltonian parameters and driving frequencies. We show that in a quench-free setting,
FDQPTs is more flexible to control than the conventional DQPTs.
We first investigate the effect of equal frequencies for the driving terms, and then show the differences with commensurate or incommensurate driving frequencies.
In particular, we demonstrate that it is possible to induce several FDQPTs within a single driving period, making the nonanalytic cusps in the return probability to form a sublattice structure in time. Moreover, we demonstrate that more FDQPTs can be observed without changing the initial state of the system \cite{Qianqian2021}. We also investigate the topological aspects of FDQPTs by computing the DTOP.

\section{Model\label{model}}

We start with a periodically driven generalized XY spin chain, whose Hamiltonian can be written as
%
{\small
\bea
\no
{\cal H}(t)=\sum_{n}\Big\{\!\!\!\!&[J-\gamma\cos(\varphi(t))]s_{n}^{x}s_{n+1}^{x}+[J+\gamma\cos(\varphi(t))]s_{n}^{y}s_{n+1}^{y}\\
\label{eq1}
&\!\!\!\!\!\!\!\!\!\!-\gamma\sin(\varphi(t))(s_{n}^{x}s_{n+1}^{y}+s_{n}^{y}s_{n+1}^{x})+h_{z}(t) s_{n}^{z}\Big\},
\eea
}
%
where, $h_{z}(t)=h_{1}+h(t)=h_{1}+h\cos(\omega t)$, and $\varphi(t)=\omega_{0}t+2\int_{0}^{t}h(t')dt'=\omega_{0}t+2(h/\omega)\sin(\omega t)$.
We choose this model as a working example to study the DQPTs in driven systems with two different
frequencies. In this paper, we concentrate on the case in which the two frequencies are commensurate with each other, such that
the Hamiltonian of the system is still periodic in time, i.e., $H(t+T_{F})=H(t)$, where $T_{F}$ is the discrete
time translational symmetry (periodicity) of the driven XY Hamiltonian in Eq. (\ref{eq1}).
Such a symmetry can be established if $\omega/\omega_{0}=p/q$ with $q,p\in{\mathbb N}$, yielding $T_{F}=2\pi q/\omega_{0}=2\pi p/\omega$.

The Hamiltonian in Eq. (\ref{eq1}) can be mapped to a free spinless fermion model
by means of the Jordan-Wigner transformation \cite{Barouch1971,LIEB1961,Jafari2012,Jafari2011,Zamani2020}
%
\bea
\no
{\cal H}(t)= \sum_{n=1}^{N} \Big[\!\!\!\!\!\!\!\!&&\Big(\frac{J}{2} c_{n}^{\dagger} c_{n+1}
-\frac{\gamma}{2} e^{-{\it i} \varphi(t)} c_{n}^{\dagger} c^{\dagger}_{n+1}+{\rm H.c.}\Big)\\
\label{eq2}
&+&h_{z}(t) (c_{n}^{\dagger} c_{n}-1/2)\Big],
\eea
%
where $c_{n}^{\dagger}, c_{n}$ are the spinless fermion creation and annihilation
operators, respectively. The Hamiltonian in Eq. (\ref{eq2}) is equivalent to
the one dimensional \textit{p}-wave superconductor with a time dependent pairing phase (magnetic flux)
$\varphi(t)$ and a periodically modulated chemical potential $h_{z}(t)$ \cite{Kitaev2001,Nakagawa2016}.

Applying Fourier transformations
%
\begin{eqnarray}
\no
c_{m} = \frac{1}{\sqrt{N}} \sum_{k} c_{k} e^{-{\it i}km},~~ c_{m}^{\dagger} = \frac{1}{\sqrt{N}} \sum_{k} c_{m}^{\dagger} e^{{\it i}km},
\end{eqnarray}
%
where the wave number $k$ is equal to $k=(2p-1)\pi/N$ and $p$ runs from $-N/2+1$ to $N/2$,
and introducing the two-component Nambu spinor $C^{\dagger}_k=(c_{k}^{\dagger},~c_{-k})$, the Hamiltonian
in Eq. (\ref{eq2}) can be decomposed as
%
\bea
\label{eq3}
{\cal H}(t)=\sum_{k}C^{\dagger}_k\mathbb{H}(k,t)C_k,
\eea
%

where
%
\bea
\label{eq4}
\mathbb{H}(k,t)=
\left(
\begin{array}{cc}
h_{z}(k,t) & {\it i}h_{xy}(k)e^{-{\it i} \varphi(t)} \\
-{\it i}h_{xy}(k)e^{{\it i} \varphi(t)}  & -h_{z}(k,t) \\
\end{array}
\right),
\eea
%

in which the parameters $h_{xy}(k)$ and $h_{z}(k,t)$ are given by
%
\bea
\label{eq5}
h_{xy}(k)  = \gamma\sin(k),~~ h_{z}(k,t)=J\cos(k)+h_{z}(t).
\eea
%

Therefore, the Bloch single particle Hamiltonian $\mathbb{H}(k,t)$ is given as
%
{\small
\bea
\label{eq6}
\mathbb{H}(k,t) = h_{xy}(k)[\sin(\varphi(t))\sigma_{x}-\cos(\varphi(t))\sigma_{y}]+h_{z}(k,t)\sigma_{z}.
\eea
}
%
The exact solution to the time-dependent Schr\"{o}dinger equation
${\it i}\frac{d}{dt}|\psi(k,t)\rangle=\mathbb{H}(k,t)|\psi(k,t)\rangle$, is found by going to the rotating
frame given by the unitary transformation $U_{R}(t)=e^{-{\it i}\varphi(t)\sigma^{z}/2}$,
%
\bea
\label{eq7}
U_{R}(t)=\left(
           \begin{array}{cc}
             e^{-{\it i}\varphi(t)/2} & 0 \\
             0 & e^{{\it i}\varphi(t)/2} \\
           \end{array}
         \right).
\eea
%
to obtain the effective time-independent Hamiltonian
%
\bea
\label{eq8}
H_{F}(k)&=&\Big[{U_R}^{\dagger}(t)\mathbb{H}(k,t)U_{R}(t)-{\it i}{U_R}^{\dagger}(t)\frac{dU_{R}(t)}{dt}\Big]\\
\no
&=&-h_{xy}(k)\sigma_{y}+B_{z}(k)\sigma_{z},
\eea
%
where, $B_{z}(k)=J\cos(k)+h_{1}-\omega_{0}/2$.

The eigenvalues and eigenvectors of the effective time-independent Hamiltonian $H_{F}(k)$ are given by
%
\bea
\label{eq9}
\varepsilon^{\pm}_{k}&=&\pm\varepsilon_{k}=\pm\sqrt{h_{xy}^{2}(k)+B_{z}^{2}(k)},\\
\no
|\chi^{\pm}_{k}\rangle&=& \frac{1}{\sqrt{f^{2}(k)+h_{xy}^{2}(k)}}
\Big[h_{xy}(k)|\mp\rangle+{\it i}f(k)|\pm\rangle\Big],
\eea
%
where $f(k)=B_{z}(k)+\varepsilon_{k}$ and $|\pm\rangle$ are eigenstates of $\sigma_z$.
Due to the decoupling of different momentum sectors, the eigenstate $|\psi(t)\rangle$ of the Hamiltonian ${\cal H}(t)$
is given by:
%
{\small
\bea
\label{eq10}
|\psi^{\pm}(t)\rangle&=&\prod_{k}|\psi^{\pm}_{k}(t)\rangle,\\
\no
|\psi^{\pm}_{k}(t)\rangle&=&U_{R}(t)e^{-{\it i}H_{F}(k)t}|\chi^{\pm}_{k}\rangle
=e^{-{\it i}\varepsilon^{\pm}_{k}t}U_{R}(t)|\chi^{\pm}_{k}\rangle.
\eea
}
%
%
\begin{figure*}[t!]
\centerline{\includegraphics[width=\linewidth]{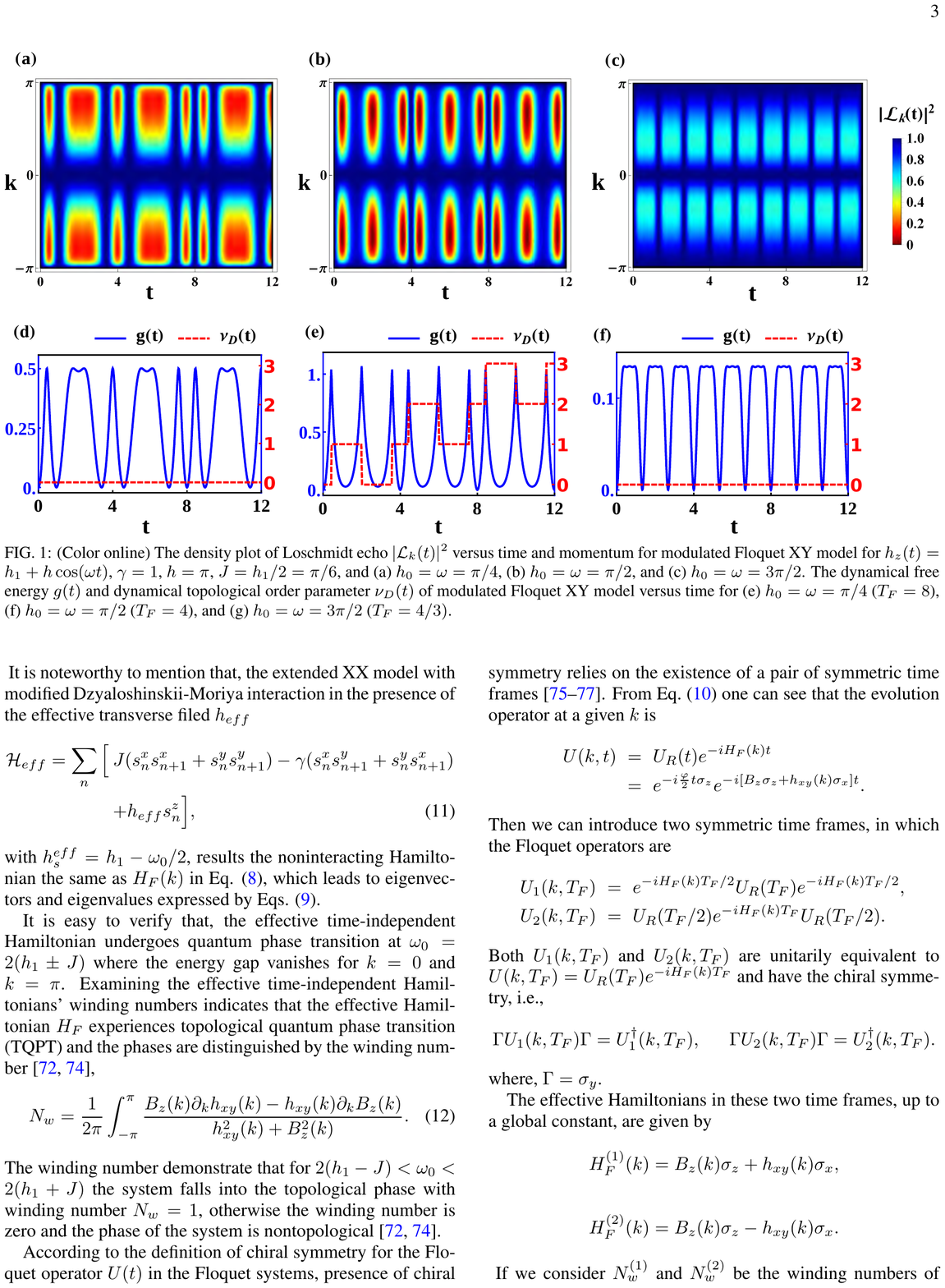}}
\centering
\caption{(Color online)
The density plot of Loschmidt echo $|{\cal L}_k (t)|^{2}$ versus time
and momentum for modulated Floquet XY model for
$\gamma=1$, $h=\pi$, $J=h_{1}/2=\pi/6$, and
(a) $\omega=\omega_{0}=\pi/4$, (b) $\omega=\omega_{0}=\pi/2$, and (c) $\omega=\omega_{0}=3\pi/2$.
The dynamical free energy $g(t)$ and dynamical topological order parameter
$\nu_{D}(t)$ of modulated Floquet XY model versus time for
(d) $\omega=\omega_{0}=\pi/4$ ($T_{F}=8$), (e) $\omega=\omega_{0}=\pi/2$ ($T_{F}=4$),
and (f) $\omega=\omega_{0}=3\pi/2$ ($T_{F}=4/3)$.}
\label{fig1}
\end{figure*}
%
It is noteworthy to mention that, the extended XX model with modified Dzyaloshinskii-Moriya interaction \cite{derzhkostatistical}
in the presence of the effective transverse field $h_{eff}$
%
\bea
\no
{\cal H}_{eff}=\sum_{n}\Big[\!\!\!\!\!\!&&J(s_{n}^{x}s_{n+1}^{x}+s_{n}^{y}s_{n+1}^{y})-\gamma(s_{n}^{x}s_{n+1}^{y}+s_{n}^{y}s_{n+1}^{x})\\
\label{eq11}
&&+h_{eff} s_{n}^{z}\Big],
\eea
%
with $h_{eff}=h_{1}-\omega_{0}/2$, results in the same noninteracting Hamiltonian
as $H_{F}(k)$ in Eq. (\ref{eq8}), yielding eigenvectors and eigenvalues as expressed by Eqs. (\ref{eq9}).

It can be verified that the effective time-independent Hamiltonian undergoes quantum phase transitions
at $\omega_{0}=2(h_{1}\pm J)$, where the energy gap closes at $k=0,\pi$.
Examining the effective Hamiltonian's winding numbers indicates that the $H_{F}$ experiences
topological quantum phase transitions (TQPTs) and the phases are distinguished by the winding number  \cite{Kitaev2001,Li2015},
%
\bea
\label{eq12}
N_{w}=\frac{1}{2\pi}\int_{-\pi}^{\pi}\frac{B_{z}(k)\partial_{k}h_{xy}(k)-h_{xy}(k)\partial_{k}B_{z}(k)}{h^{2}_{xy}(k)+B^{2}_{z}(k)}.
\eea
%
The winding number demonstrates that for $2(h_{1}-J)<\omega_{0}<2(h_{1}+J)$ the system falls into the topological phase with winding number $N_{w}=1$,
otherwise the winding number is zero and the phase of the system is non-topological \cite{Kitaev2001,Li2015}.

In Floquet systems, the presence of chiral symmetry for the Floquet operator relies on the existence of a pair of symmetric time frames \cite{Asboth2012,Asboth2013,Longwen2019PRB}.
From Eq.~(\ref{eq10}), one can see that the evolution operator at a given $k$ is
%
\bea
\no
U(k,t)&=&U_{R}(t)e^{-iH_{F}(k)t}\\
\no
&=&e^{-i\frac{\varphi}{2}t\sigma_{z}}e^{-{\it i}[B_{z}\sigma_{z}+h_{xy}(k)\sigma_{x}] t}.
\eea
%
We can then introduce two symmetric time frames, in which the Floquet operators are
%
\bea
\no
U_{1}(k,T_{F})&=& e^{-{\it i}H_{F}(k)T_{F}/2}U_{R}(T_{F})e^{-{\it i}H_{F}(k)T_{F}/2},\\
\no
U_{2}(k,T_{F})&=& U_{R}(T_{F}/2)e^{-{\it i}H_{F}(k)T_{F}}U_{R}(T_{F}/2).
\eea
%
Both $U_{1}(k,T_{F})$ and $U_{2}(k,T_{F})$ are unitarily equivalent to $U(k,T_{F})=U_{R}(T_{F})e^{-iH_{F}(k)T_{F}}$. They share the chiral symmetry, i.e.,
%
\begin{equation}
\no
\Gamma U_{1}(k,T_{F})\Gamma=U_{1}^{\dagger}(k,T_{F}),~~~~~~\Gamma U_{2}(k,T_{F})\Gamma=U_{2}^{\dagger}(k,T_{F}),
\end{equation}
%
where $\Gamma=\sigma_{x}$.

The effective Hamiltonians in these two time frames, up to a global constant, are given by
%
\begin{equation}
\no
H_{F}^{(1)}(k)=B_{z}(k)\sigma_{z}-h_{xy}(k)\sigma_{y},
\end{equation}
%
%
\begin{equation}
\no
H_{F}^{(2)}(k)=B_{z}(k)\sigma_{z}+h_{xy}(k)\sigma_{y}.
\end{equation}
%
%
\begin{figure*}[t!]
\centerline{\includegraphics[width=\linewidth]{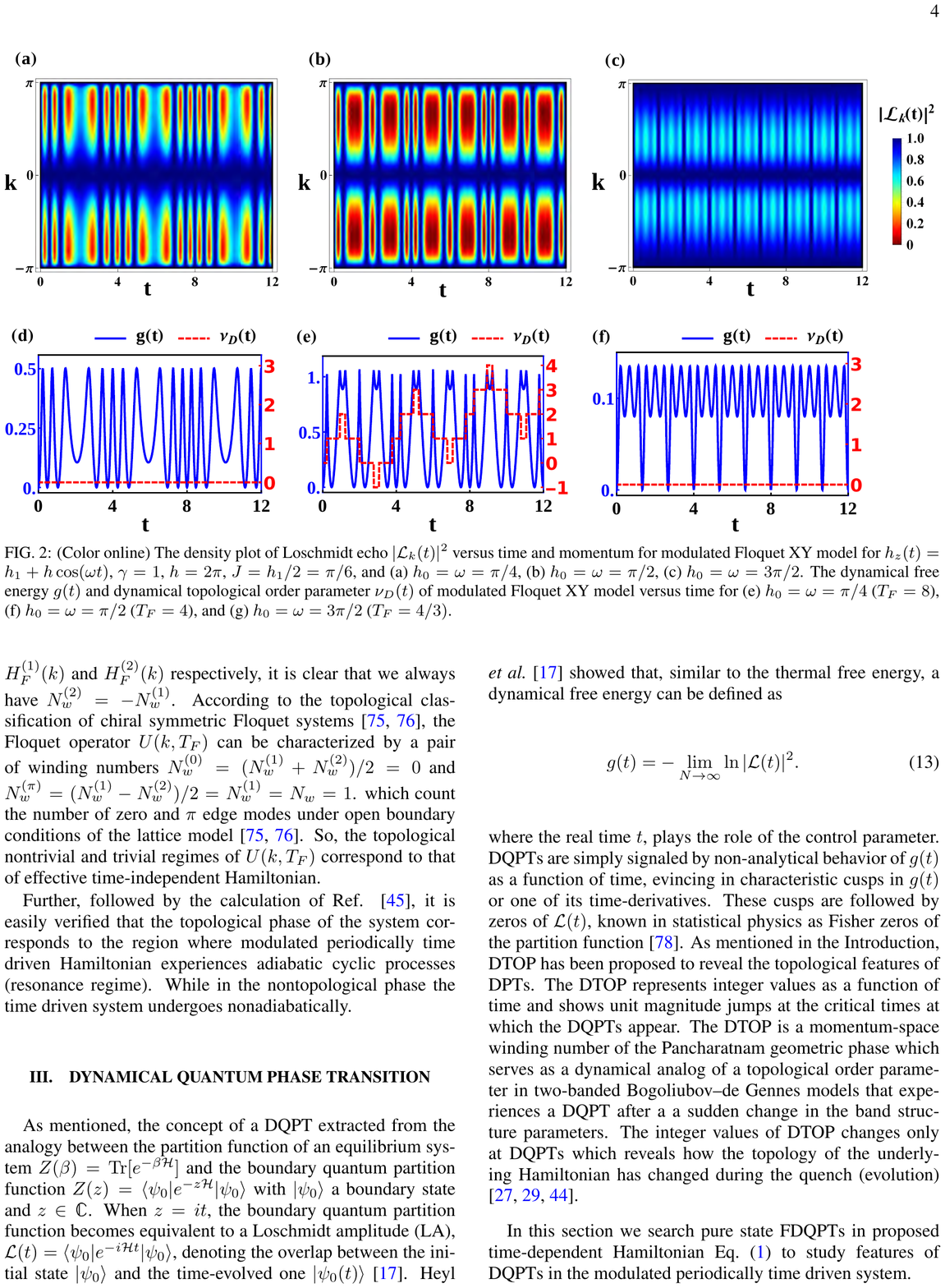}}
\centering
\caption{(Color online)
The density plot of Loschmidt echo $|{\cal L}_k (t)|^{2}$ versus time
and momentum for modulated Floquet XY model for
$\gamma=1$, $h=2\pi$, $J=h_{1}/2=\pi/6$, and
(a) $\omega=\omega_{0}=\pi/4$, (b) $\omega=\omega_{0}=\pi/2$, (c) $\omega=\omega_{0}=3\pi/2$.
The dynamical free energy $g(t)$ and dynamical topological order parameter
$\nu_{D}(t)$ of modulated Floquet XY model versus time for
(d) $\omega=\omega_{0}=\pi/4$ ($T_{F}=8$), (e) $\omega=\omega_{0}=\pi/2$ ($T_{F}=4$),
and (f) $\omega=\omega_{0}=3\pi/2$ ($T_{F}=4/3)$.}
\label{fig2}
\end{figure*}
%
If we consider $N^{(1)}_{w}$ and $N^{(2)}_{w}$ as the winding numbers of $H_{F}^{(1)}(k)$ and $H_{F}^{(2)}(k)$ respectively,
it is clear that we always have $N^{(2)}_{w}=-N^{(1)}_{w}$. According to the topological classification
of chiral symmetric Floquet systems~\cite{Asboth2012,Asboth2013}, the Floquet operator $U(k,T_{F})$ can be characterized
by a pair of winding numbers $N^{(0)}_{w}=(N^{(1)}_{w}+N^{(2)}_{w})/2=0$ and $N^{(\pi)}_{w}=(N^{(1)}_{w}-N^{(2)}_{w})/2=N^{(1)}_{w}=N_{w}=1$.
which count the number of zero and $\pi$ edge modes under open boundary conditions of the lattice model~\cite{Asboth2012,Asboth2013}.
So, the topological nontrivial and trivial regimes of $U(k,T_F)$ correspond to that of effective time-independent Hamiltonian.

Furthermore, following the calculations of Ref. \cite{Zamani2020}, it is easy to verify that the topological phase of
the system corresponds to the region where the periodically modulated Hamiltonian experiences adiabatic cyclic
processes (resonance regime). While in the nontopological phase, the driven system undergoes nonadiabatic evolution.

\section{Dynamical quantum phase transition}

As mentioned above, the concept of DQPT is extracted from the analogy between the partition function of an equilibrium system
$Z(\beta)={\rm Tr}(e^{-\beta {\cal H}})$ and the boundary partition function $Z(z)=\langle\psi_{0}|e^{-z {\cal H}}|\psi_{0}\rangle$,
with $|\psi_{0}\rangle$ being a boundary state and $z \in \mathbb{C}$.
When $z=it$, the boundary partition function becomes equivalent to the  Loschmidt amplitude (LA),
${\cal L}(t)=\langle\psi_{0}|e^{-{\it i}  {\cal H}t}|\psi_{0}\rangle$, denoting the overlap between the initial state $|\psi_{0}\rangle$
and the time-evolved one $|\psi_{0}(t)\rangle$~\cite{Heyl2013}.
Heyl {\em et al.}~\cite{Heyl2013} showed  that, similar to the thermal free energy, a dynamical free energy can be defined as
%
\bea
\label{eq13}
g(t)=-\frac{1}{N}\lim_{N\rightarrow\infty} \ln|{\cal L}(t)|^{2}.
\eea
%
where the real time $t$ plays the role of the control parameter.
DQPTs are simply signalled by the non-analytical behavior of $g(t)$ as a function of time,
evincing in characteristic cusps in $g(t)$ or one of its time-derivatives.
These cusps are followed by zeros of ${\cal L}(t)$, known in statistical physics as Fisher zeros of the
partition function~\cite{Fisher1967}.
As mentioned in the Introduction, DTOP has been proposed to reveal the topological features of DQPTs. The DTOP takes
integer values as a function of time and shows unit magnitude jumps at the critical times when the DQPT
happens.

In this section we investigate FDQPTs for both pure and mixed Floquet states in the system described by time-dependent Hamiltonian Eq.~(\ref{eq1}).
The focus of our study are the control of FDQPTs in periodically modulated systems.

%
\begin{figure*}[t]
\centerline{\includegraphics[width=\linewidth]{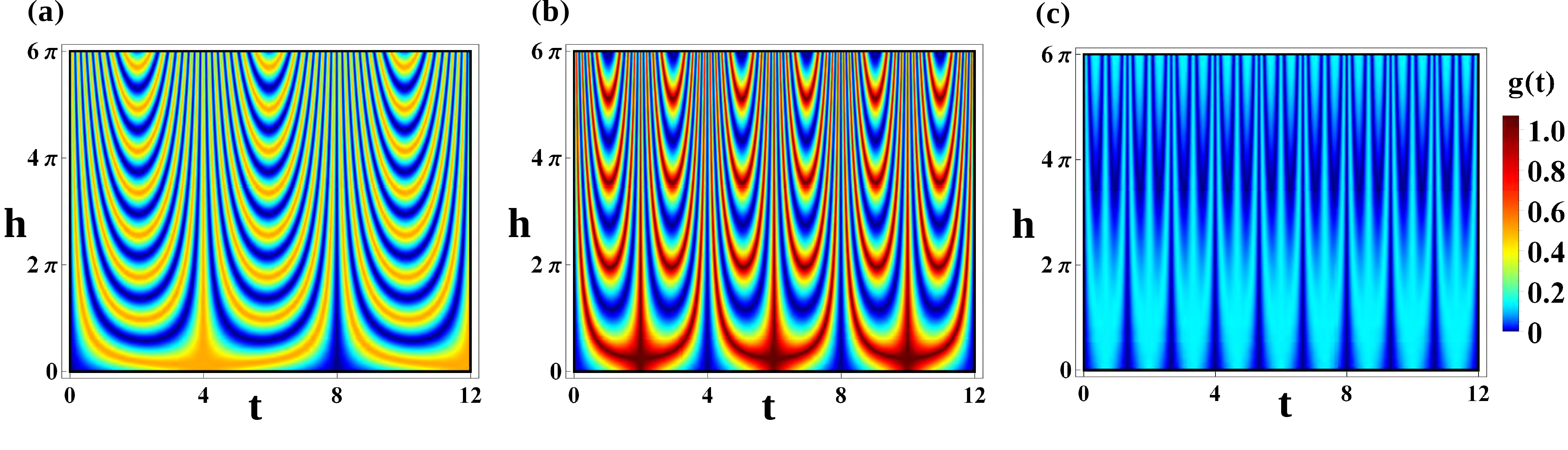}}
\centering
\caption{(Color online)
The density plot of dynamical free energy $g(t)$ versus time
and $h$ for modulated Floquet XY model for
$\gamma=1$, $J=h_{1}/2=\pi/6$, and
(a) $\omega=\omega_{0}=\pi/4$ $(T_{F}=8)$, (b) $\omega=\omega_{0}=\pi/2$ $(T_{F}=4)$,
(c) $\omega=\omega_{0}=3\pi/2$ $(T_{F}=4/3)$.}
\label{fig3}
\end{figure*}
%

\subsection{Pure state FDQPTs}
According to Eq. (\ref{eq10}), the initial and time evolved ground states of the original Hamiltonian are expressed by:

%
\bea
\no
|\psi^{-}(t)\rangle&=&\prod_{k}|\psi^{-}_{k}(t)\rangle=
\prod_{k} e^{-{\it i}\varepsilon^{-}_{k}t}U_{R}(t)|\chi^{-}_{k}\rangle,\\
\label{eq14}
|\psi^{-}(0)\rangle&=&\prod_{k}|\chi^{-}_{k}\rangle
\eea
%

It is straightforward to show that the return probability--Loschmidt echo-- to
the ground state of the proposed Floquet model is given by
%
\bea
\label{eq15}
{\cal L}(t)&=&\langle\psi^{-}(0)|\psi^{-}(t)\rangle=\prod_{k}{\cal L}(k,t),\\
\no
{\cal L}(k,t)&=&\langle\chi^{-}_{k}|\psi^{-}_{k}(t)\rangle=e^{-{\it i}\varepsilon^{-}_{k}t}\langle\chi^{-}_{k}|U_{R}(t)|\chi^{-}_{k}\rangle,\\
\no
&=&e^{-i\varepsilon^{-}_{k}t}e^{-{\it i}\varphi(t)/2}\frac{f^{2}(k)
+e^{{\it i}\varphi(t)}h^{2}_{xy}(k)}{f^{2}(k)+h^{2}_{xy}(k)},
\eea
%
The DQPTs occur at the time instances at which at least one factor in LA becomes zero, i.e., ${\cal L}_{k^{\ast}}(t^{\ast})=0$.
Referring to Eq. (\ref{eq15}), we find that FDQPT happens only when there is a mode $k^{\ast}$, which satisfies $J\cos(k)+h_{1}-\omega_{0}/2=0$,
which leads to
%
\bea
\label{eq:FDQPT}
2(h_{1}-J)<\omega_{0}<2(h_{1}+J),
\eea
%

at time instances $t^{\ast}$, when the equation
%
\bea
\label{eq16}
\omega_{0} t^{\ast}+2\frac{h}{\omega}\sin(\omega t^{\ast})=(2n+1)\pi,
\eea
%
is fulfilled. The condition Eq. (\ref{eq:FDQPT}), reveals that FDQPTs appear if the effective time-independent Hamiltonian
in Eq. (\ref{eq8}) is topologically nontrivial which is controlled by the driving frequency $\omega_{0}$.
However, the time scale of FDQPTs is controlled by both the driving frequency $\omega$, and $\omega_{0}$
and Hamiltonian parameter $h$. Accordingly, these properties make it possible to easily engineer and control the FDQPTs.
Although we focus on commensurate case, i.e., $\omega/\omega_{0}=p/q$ with $p, q\in{\mathbb N}$,
the equations obtained above are valid also for the cases with $\omega/\omega_{0}\neq p/q$.

To understand the effect of Hamiltonian parameter $h$ on FDQPTs, in this section we consider
$\omega=\omega_{0}$,  ($p=q$) case. In such a case, the equation of real-time nonanalyticity reduces to
%
\bea
\label{eq17}
\omega_{0} t^{\ast}+2\frac{h}{\omega_{0}}\sin(\omega_{0} t^{\ast})=(2n+1)\pi.
\eea
%

A purely analytical solution to Eq. (\ref{eq16}) or (\ref{eq17}) is not tractable, which requires numerical solutions.
Nevertheless, it can be
verified that  Eq. (\ref{eq17}) is satisfied by
%
\bea
\label{eq18}
t_{m,F}^{\ast}=(2m+1)\frac{T_{F}}{2},~~m\in{\mathbb N},
\eea
%
which is the only solution for $h=0$ although other numerical solutions show up for $h\neq0$.
Further, it can be easily shown that Eq. (\ref{eq17}) is preserved under the transformation $t^{\ast}\rightarrow t^{\ast}+T_{F}$,
which means that the time-periodicity of FDQPTs is the same as that of the Floquet Hamiltonian.
Moreover, the equation of real-time nonanalyticity is preserved under the transformation $t^{\ast}\rightarrow T_{F}-t^{\ast}$,
which means that the patterns of FDQPTs and dynamical free energy have the reflection symmetry with respect to $t=T_{F}/2$
within each driving period. We should mention that, the Hamiltonian in Eq. (\ref{eq1}) reduces to the Hamiltonian of Refs. \cite{Zamani2020,yang2019floquet} if  we remove the time dependence of the transverse magnetic field, putting $h=0$. In our model, for $\omega=\omega_{0}$, the transverse field in the Hamiltonian, Eq. (\ref{eq1}), is still time dependent while in Refs. \cite{Zamani2020,yang2019floquet} the transverse field is time-independent.

The density plot of the Loschmidt echo $|{\cal L}(k,t)|^{2}$ and the dynamical free energy $g(t)$ have been displayed
for $J=h_1/2=\pi/6$, and $\omega_{0}=\omega$ in Figs. (\ref{fig1})(a)-(f) for $h=\pi$, and in Figs. (\ref{fig2})(a)-(f)
for  $h=2\pi$. It can be clearly seen that, in the region where the time-independent Hamiltonian $H_{F}$ is topological
and the system is in resonance regime (Fig. \ref{fig1}(b) and Fig. \ref{fig2}(b)), there exist critical
points $k^{\ast}$ and $t^{\ast}$ where ${\cal L}_{k^{\ast}}(t^{\ast})$ becomes zero. Interestingly, there are no such
critical points in non resonance regime (Figs. \ref{fig1}(a), \ref{fig1}(c), \ref{fig2}(a) and \ref{fig2}(c)).
Consequently, the nonanalyticity in the dynamical free energy and FDQPTs occur for the driving frequency, at which the
system is in topological phase. Cusps in $g(t)$ in Figs.~\ref{fig1}(e)-\ref{fig2}(e) are clearly visible, implying
FDQPTs. For the driving frequency at which the system is in non-topological phase, the dynamical
free energy shows completely analytic and smooth behavior (Figs.~\ref{fig1}(a)-\ref{fig1}(c)-\ref{fig2}(a)-\ref{fig2}(c)).
Moreover, as is clearly seen in Figs.~\ref{fig1}(e)-\ref{fig2}(e) that the pattern of FDQPTs has reflection symmetry with respect to
$t=T_{F}/2$ within each driving period, as expected from Eq. (\ref{eq17}).\\

It should be mentioned that $t_{m,F}^{\ast}$ is independent of the value of $h$,
which makes $t_{m,F}^{\ast}$ to be the only time scale of FDQPT
for $h=0$ \cite{Zamani2020,Jafari2021,yang2019floquet}.
In other words, the dynamical phase transition for $h=0$ takes place only once within every Floquet time period
and its periodicity is the same as that of the Floquet Hamiltonian \cite{Zamani2020,Jafari2021,yang2019floquet}.
While  for $h\neq0$, FDQPT time scales (solutions of Eq. (\ref{eq17})) are not periodic within each Floquet time period, the global pattern of FDQPTs
is repeated at every driving period (Fig. \ref{fig1}(b) and Fig. \ref{fig2}(b)). The density plot of $g(t)$ versus time and $h$
has been shown in Figs. \ref{fig3}(a)-(f).
%
\begin{figure*}[t]
\begin{minipage}{\linewidth}
\centerline{\includegraphics[width=\linewidth]{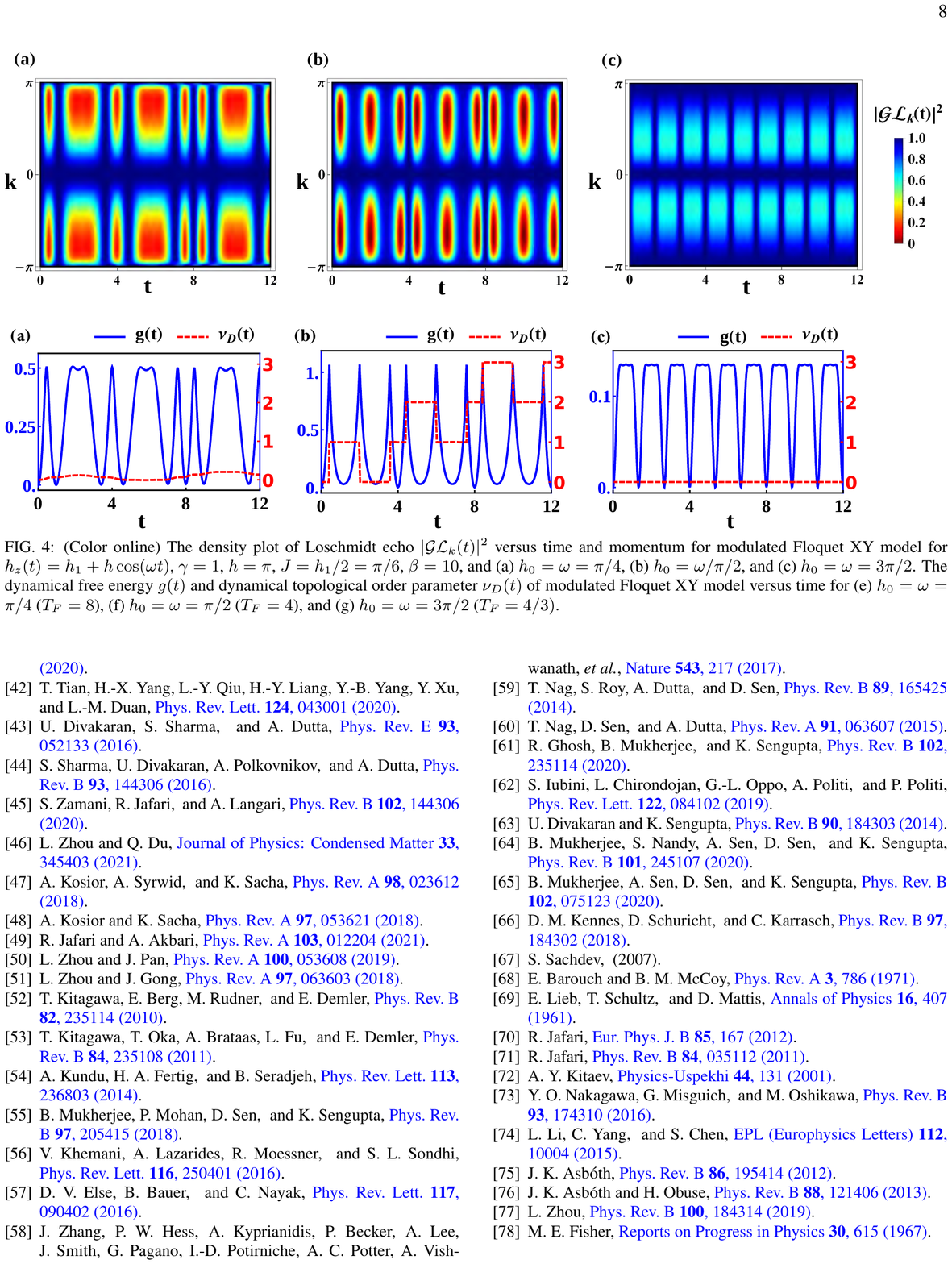}}
\centering
\end{minipage}
\caption{(Color online)
The mixed state dynamical free energy $g(t)$ and dynamical topological order parameter
$\nu_{D}(t)$ of modulated Floquet XY model versus time for
$\gamma=1$, $h=\pi$, $J=h_{1}/2=\pi/6$, $\beta=10$, and
(a) $\omega=\omega_{0}=\pi/4$ ($T_{F}=8$), (b) $\omega=\omega_{0}=\pi/2$ ($T_{F}=4$),
and (c) $\omega=\omega_{0}=3\pi/2$ ($T_{F}=4/3)$.}
\label{fig4}
\end{figure*}
%
%
\begin{figure*}[t]
\centerline{\includegraphics[width=\linewidth]{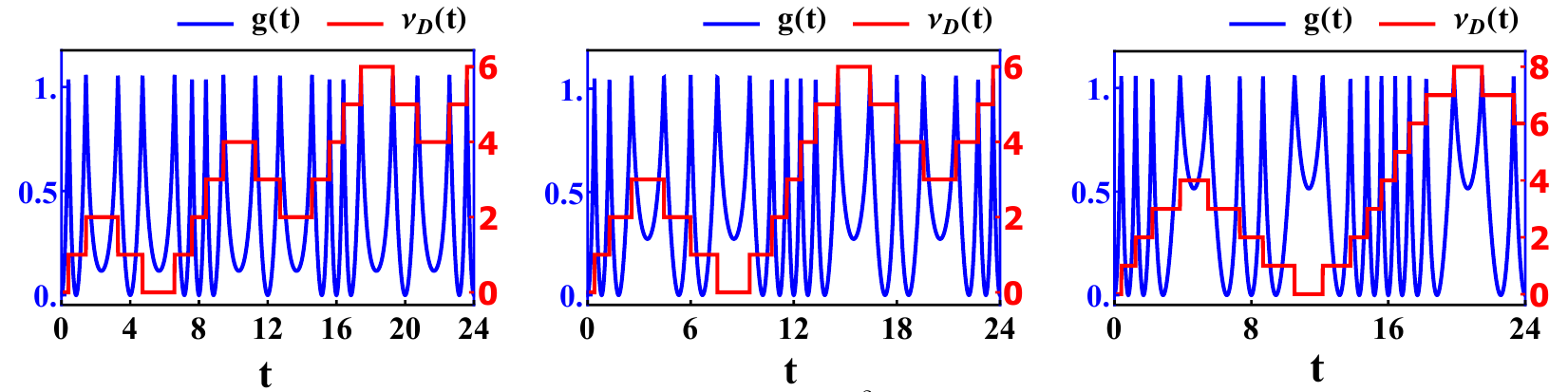}}
\centering
\caption{(Color online)
The dynamical free energy $g(t)$ and dynamical topological order parameter
$\nu_{D}(t)$ of modulated Floquet XY model versus time for
$\gamma=1$, $h=\pi$, $J=h_{1}/2=\pi/6$, $\omega_{0}=\pi/2$ and
(a) $\omega=\omega_{0}/2=\pi/4$ ($T_{F}=8$), (b) $\omega=\omega_{0}/3=\pi/6$ ($T_{F}=12$),
and (c) $\omega=\omega_{0}/4=\pi/8$ ($T_{F}=32)$.}
\label{fig5}
\end{figure*}
%
We see that the system undergoes more and more FDQPTs within a single driving period with the increase
of $h$, with a cusp observed at every critical time $t^{\ast}$ as predicted
precisely by Eq. (\ref{eq17}).
In summary,  the Hamiltonian parameter $h$ induces several FDQPTs within a single driving
period, while the FDQPT time scale ($t^{\ast}=T_{F}/2$) is fixed for $h=0$.

\subsubsection{Dynamical topological order parameter}
As mentioned above, analogous to order parameters at equilibrium quantum phase transitions, a dynamical topological order parameter
is proposed to capture DQPTs~\cite{budich2016dynamical,Bhattacharya}.
The DTOP is quantized and its unit magnitude jump at the time of DQPT reveals the topological feature of DQPT~\cite{budich2016dynamical,Bhattacharya,Sharma2016}.
This dynamical topological order parameter is extracted from the “gauge-invariant” Pancharatnam geometric
phase associated with the Loschmidt amplitude~\cite{budich2016dynamical,Bhattacharya}.
In other words, the DTOP is a momentum-space winding number of the Pancharatnam geometric phase, which serves as a dynamical
analogue of a topological order parameter in two-band Bogoliubov–de Gennes models that experiences a DQPT after
a sudden change in the band structure parameters. The integer values of DTOP changes only at DQPTs, which reveals
how the topology of the underlying Hamiltonian has changed during the quench \cite{budich2016dynamical,Bhattacharya,Sharma2016}.
The dynamical topological order parameter is defined as~\cite{budich2016dynamical}
%
\begin{eqnarray}
\label{eq19}
\nu_D(t)=\frac{1}{2\pi}\int_0^\pi\frac{\partial\phi^G(k,t)}{\partial k}\mathrm{d}k,
\end{eqnarray}
%
where the geometric phase $\phi^G(k,t)$ is obtained by subtracting the dynamical
phase $\phi^{D}(k,t)$ from the total phase $\phi(k,t)$, i.e., $\phi^G(k,t)=\phi(k,t)-\phi^{D}(k,t)$.

The total phase $\phi(k,t)$ is the phase factor of return amplitude,
i.e., ${\cal L}_{k}(t)=|{\cal L}_{k}(t)|e^{i\phi(k,t)},$ and $\phi^{D}(k,t)=-\int_0^t \langle \psi_{k}^{-}(t')|H(k,t')|\psi_{k}^{-}(t')\rangle dt',$
in which $\phi(k,t)$ and $\phi^{D}(k,t)$ can be calculated  as follows
%
\begin{eqnarray}
\label{eq20}
\phi(k,t)&=& -(\varepsilon^{-}_{k}t+\varphi(t)/2)\\
\no
&+& \tan^{-1}\Big(\frac{f^{2}(k)\sin(\varphi(t))}
{h_{xy}^{2}(k) + f^{2}(k) \cos(\varphi(t))}\Big),
\end{eqnarray}
%
%
%
{\small
\begin{eqnarray}
\label{eq21}
\phi^{D}(k,t)=
-\varepsilon^{-}_{k}t+\Big[\frac{f^{2}(k)-h^{2}_{xy}(k)}{2(f^{2}(k)+h^{2}_{xy}(k))}\Big](\varphi(t)-\varphi(0)).
\end{eqnarray}
}
%
%
The DTOP of our model has been plotted for $\omega=\omega_{0}$ case in Figs.~\ref{fig1}(d)-(f) and \ref{fig2}(d)-(f) for different values of driven frequency and Hamiltonian parameters in resonance and non-resonance regimes. As seen, the DTOP is zero when FDQPTs are absent while
the DTOP displays singular changes in successive critical times $t_{n}^{\ast}$ in the adiabatic resonance regime where FDQPTs occur.
The unit jumps in $\nu_{D}(t)$ feature the topological aspects of DQPTs, where the phase of time-independent Floquet Hamiltonian
$H_{F}$ is topological. While the DTOP for $h\neq0$ shows unit jumps up and down at $t^{\ast}_{n}$,
the DTOP layout within every Floquet time period is preserved up to a unit jumps up (Figs.~\ref{fig1}(e) and \ref{fig2}(e))
at each driving period. We should mention that for $h=0$, DTOP shows the unit jumps up at $t^{\ast}=T_{F}/2$ \cite{Zamani2020,Jafari2021,yang2019floquet}.

\subsection{Mixed state FDQPTs}
In far-from-equilibrium experiments~\cite{flaschner2018observation,jurcevic2017direct},
the initial state in which system is prepared is generally not a pure state but rather a mixed state. This leads us to introduce
generalized Loschmidt amplitude (GLA) for mixed thermal states, which perfectly reproduces the nonanalyticity appear
in the DQPTs of pure states \cite{Heyl2017,Bhattacharya}. Here we investigate the notion of mixed state FDQPTs in Floquet dynamics,
governed by Eq. (\ref{eq1}). The GLA for thermal mixed state is given as
%
\begin{equation}
{\cal GL}(t) =\prod_{k} {\cal GL}_{k}(t)=\prod_{k}Tr \Big(\rho_{k}(0) U(t)\Big),
\label{eq22}
\end{equation}
%
where $\rho_{k}(0)$ is the mixed state density matrix at time $t=0$, and $U(t)$ is the time-evolution operator.
By a rather lengthy calculation, one can obtain an exact expression for GLA (See Appendix \ref{APA})
%
\bea
\label{eq23}
{\cal GL}_{k}(t)={\cal R}(k,t)+{\it i}\,{\cal I}(k,t)\tanh(\beta\varepsilon_{k})
\eea
%

where
%
{\small
\bea
\no
{\cal R}(k,t)&=&\cos(\varepsilon_{k}t)\cos(\varphi(t)/2)-\frac{B_{z}(k)}{\varepsilon_{k}}\sin(\varepsilon_{k}t)\sin(\varphi(t)/2),\\
\no
{\cal I}(k,t)&=&\sin(\varepsilon_{k}t)\cos(\varphi(t)/2)+\frac{B_{z}(k)}{\varepsilon_{k}}\cos(\varepsilon_{k}t)\sin(\varphi(t)/2).
\eea
}
%
The dynamical free energy of generalized Loschmidt echo $g(t)$ has been displayed versus time $t$ and $k$ in Figs.~\ref{fig4}(a)-(c) for $\omega=\omega_{0}$ case, for different values of driving frequency at $\beta=10$. As is clear from the figures, the nonanalyticity in the dynamical free energy of GLA appear in the resonance regime (Fig. \ref{fig4}(b)) and correctly reproduces the critical time $t^{\ast}$ observed during the pure state FDQPT. It should be mentioned that for temperatures higher than the temperature associated with the minimum energy gap of the time independent Hamiltonian, the finger print of DQPT are washed out \cite{Zamani2020,Bhattacharya}.

Analogous to the pure state FDQPT, topological invariant has also been established for mixed state DQPT to reveal its topological feature~\cite{Bhattacharya}. In the mixed state DQPT, the total phase and dynamical phase are defined as $\phi(k,\beta,t)={\rm Arg}\Big[Tr\big(\rho(k,\beta,0)U(t)\big)\Big]$, and $\phi^{D}(k,\beta,t)=-\int_{0}^{t} \tr[\rho(k,\beta,t')H(k,t')]dt'$, respectively.
The topological invariant $\nu_D(t)$ for mixed states can then be obtained using Eq.~(\ref{eq19}) in which $\phi^{G}(k,\beta,t)=\phi(k,\beta,t)-\phi^{D}(k,\beta,t)$. A rather lengthy calculation results in exact expressions for
$\phi(k,\beta,t)$ and $\phi^{D}(k,\beta,t)$ for a mixed state (see Appendix \ref{APA}).\\

The mixed state DTOP has been illustrated in Fig. \ref{fig4} at $\beta=10$ for $\omega=\omega_{0}$ case, for different values of driving frequencies.
One can clearly see that $\nu_D(t)$ exhibits a perfect quantization (unit jump) as a function of time
between two successive critical times $t^{\ast}$ in the resonance regime, as shown in Fig. \ref{fig4}(b), while it is zero in no-FDQPTs
regime (Figs. \ref{fig4}(a)-\ref{fig4}(c)). The quantized structure of $\nu_D(t)$ is only observed as far as temperatures
are smaller than the temperature associated with the minimum energy gap of the time independent Hamiltonian~\cite{Bhattacharya,Zamani2020}.

\subsection{Pure state FDQPTs for $\omega/\omega_{0}=p/q$}
To better understand the effect of second driven frequency $\omega$ on FDQPTs, here, we investigate the case of
$\omega/\omega_{0}=p/q$ for a constant $h=\pi$. In such cases, the discrete time translational
symmetry of the modulated Floquet Hamiltonian (Eq. (\ref{eq1})) is given by $T_F=2\pi q/\omega_{0}=2\pi p/\omega$,
and the equation of real-time nonanalyticity  specifying the time scale of FDQPTs is given by Eq. (\ref{eq16}).
Further, it can be shown that Eq. (\ref{eq16}) is preserved under the transformation $t^{\ast}\rightarrow t^{\ast}+T_{F}$
and $t^{\ast}\rightarrow T_{F}-t^{\ast}$, which means that the time period of FDQPT is the same as that of the Floquet Hamiltonian,
and the dynamical free energy has the reflection symmetry with respect to $t=T_{F}/2$ within each driven period.

In addition, we can show that the condition of real-time nonanalyticity in Eq. (\ref{eq16}) is satisfied at
%
\bea
\label{eq24}
t_{m,F}^{\ast}=(2m+1)\frac{T_{F}}{2},~~m\in{\mathbb N}; T_F=2\pi q/\omega_{0}=2\pi p/\omega,
\eea
%
where $q$ is an odd number. This means that the time scale $t^{\ast}=T_{F}/2$ at which the system shows FDQPT for $h=0$ \cite{Zamani2020,Jafari2021,yang2019floquet} is still the FDQPT time scale for $h\neq0$.
However, for even  $q$, the system does not show FDQPT at $t=T_{F}/2$ and $h\neq0$.
So the FDQPT time scale ($t^{\ast}=T_{F}/2$) for $h\neq0$ can be controlled by the ratio of the two driving frequencies  ($\omega_{0}/\omega=p/q$).

The dynamical free energy $g(t)$ of the model has been plotted in the region where the system undergoes FDQPTs
for  $p=1$ and different values of  $q$ in Figs. (\ref{fig5})(a)-(c) at $h=\pi$. As seen, the Floquet time period and
the number of FDQPTs in a single driving period can be raised by increasing the ratio of $\omega_{0}/\omega=q/p$.
As expected, for  $q=2,4$ (Figs. (\ref{fig5})(a)-(\ref{fig5})(c)) the dynamical free energy does not show nonanalyticity at
$t=T_{F}/2$, while for  $q=3$ (Fig. (\ref{fig5})(b)) the cusp at $t=T_{F}/2$ represents the FDQPT.
Moreover, as is clearly seen, the dynamical free energy has the reflection symmetry with respect to $t=T_{F}/2$ within each driving
period. We have also plotted the DTOP for  $p=1$  and different values of  $q$ in Figs. (\ref{fig5})(a)-(c), which shows the quantized jump whenever
a Floquet DQPT happens.

\section{Conclusion}
In this work, we introduced a quench-free route to engineer and control FDQPTs. The key idea of our strategy is to apply two driving fields with commensurate frequencies to a system. The first field guides the periodic Floquet dynamics, whereas the second field with a higher frequency controls FDQPTs within each period of the first drive. Our approach is demonstrated in a driven XY spin chain, where we observe rich patterns of FDQPTs within each driving period for both pure and mixed initial states. These transitions are further characterized by quantized jumps of DTOPs. Our discovery unveiled the flexibility of Floquet systems in the engineering and control of DQPTs compared with the conventional cases following a single quench. Therefore, the Floquet system with multiple driving frequencies can work as a useful dynamical platform to engineer and control phase transitions out-of-equilibrium.
It is worthwhile to mention that when deviating slightly from the commensurate to incommensurate cases the DQPTs are still present but the critical time
is not the same as that of the commensurate case and the periodicity of the dynamical free energy is wiped out (for more detail see appendix \ref{APB}). Hence, in the case of incommensurate case the DQPTs are not periodic in time as well as the Hamiltonian.

Moreover, we would like to mention that our findings may be verified experimentally by a negatively charged nitrogen-vacancy center by which the non-interacting single mode Hamiltonian $\mathbb{H}(k,t)$, in two-band insulator can be simulated experimentally \cite{Ma}.

\section*{Acknowledgements}
R. J is grateful to Henrik Johannesson for reading the manuscript and valuable comments.

\appendix

\section{Dynamical phase transition for mixed state\label{APA}}

The mixed state density matrix at time $t=0$ describing the system  at  thermal equilibrium with a bath corresponding to the initial Hamiltonian
$H_{F}(k)=q(k)\mathbb{1} + \vec{h}_{{\it l}}(k)\cdot {\vec{\sigma}}$ can be written as \cite{Heyl2017,Bhattacharya}
%
\bea
\label{eqA1}
\rho_{F} (k,0)=\frac{e^{-\beta H_{F}(k)}}{{\rm Tr}(e^{-\beta H_{F}(k)})}
=\frac{1}{2}(\mathbb{1}-\Delta{{\hat n_{\it l}}}(k)\cdot {\vec {\sigma}})
\eea
%
where $\beta$ is the inverse temperature,
$ \Delta=\tanh(\beta |\vec{h}_{{\it l}}(k)|)$, $\hat{n}_{{\it l}}(k) = \vec{h_{\it l}}(k)/|\vec{h}_{{\it l}}(k)|$.\\

The mixed state density matrix at time $t$ of the time-dependent Hamiltonian $\mathbb{H}(k,t)$
is given by
%
\bea
\label{eqA2}
{\rho}(k,t)= U_{F}(t) \rho_{F} (k,0) U_{F}^{\dagger}(t),
\eea
%
where $U_{F}(t)=U_{R}(t)e^{-{\it i} H_{F}(k)t}$.

The generalised Loschmidt overlap amplitude (GLOA) for each $k$ mode is defined as \cite{Heyl2017,Bhattacharya}:
%
\bea
\label{eqA3}
{\cal  GL}(t)&=&\prod_{k} {\cal GL}_{k}(t),\\
\no
{\cal GL}_{k}(t)&=&Tr \Big(\rho(k,0) U_{F}(t)\Big)\\
\no
&=&{\cal R}(k,t)+{\it i}\,{\cal I}(k,t)\tanh(\beta\varepsilon_{k})
\eea
%

where
%
{\small
\bea
\no
{\cal R}(k,t)&=&\cos(\varepsilon_{k}t)\cos(\varphi(t)/2)-\frac{B_{z}(k)}{\varepsilon_{k}}\sin(\varepsilon_{k}t)\sin(\varphi(t)/2)\\
\no
{\cal I}(k,t)&=&\sin(\varepsilon_{k}t)\cos(\varphi(t)/2)+\frac{B_{z}(k)}{\varepsilon_{k}}\cos(\varepsilon_{k}t)\sin(\varphi(t)/2).
\eea
}
%

Moreover, for mixed state DQPT topological invariant has been proposed to lay out its topological characteristics\cite{Heyl2017,Bhattacharya}.
In the mixed state DQPT the total phase and dynamical phase are given as
$$\phi(k,\beta,t)={\rm Arg}\Big[{\rm Tr}\Big[\rho(k,\beta,0)U(t)\Big]\Big];$$
and
$$\phi^{D}(k,\beta,t)=-\int_{0}^{t} {\rm Tr}\Big[\rho(k,\beta,t')H(k,t')\Big]dt',$$
respectively.
The topological invariant $\nu_D(t)$ can be calculated using Eq.~(\ref{eq21}) for mixed state in which
$$\phi^{G}(k,\beta,t)=\phi(k,\beta,t)-\phi^{D}(k,\beta,t).$$
After a lengthy calculation, one can obtain the total phase $\phi(k,\beta,t)$ and the dynamical phase $\phi^{D} (k,\beta,t)$ as follows
%
\begin{eqnarray}
\label{eqA4}
\phi(k,\beta,t)&=&\arctan[\frac{{\cal R}(k,t)}{{\cal I}(k,t)}\tanh(\beta\varepsilon_{k})]\\
\no
\phi^{D} (k,\beta,t)&=&\tanh(\beta\varepsilon_{k})\Big[-\varepsilon^{-}_{k}t-\frac{B_{z}(k)}{2\varepsilon_{k}}(\varphi(t)-\varphi(0))\Big].
\end{eqnarray}
%

\section{Dynamical phase transition: incommensurate case\label{APB}}

%
\begin{figure}[t]
\centerline{\includegraphics[width=\linewidth]{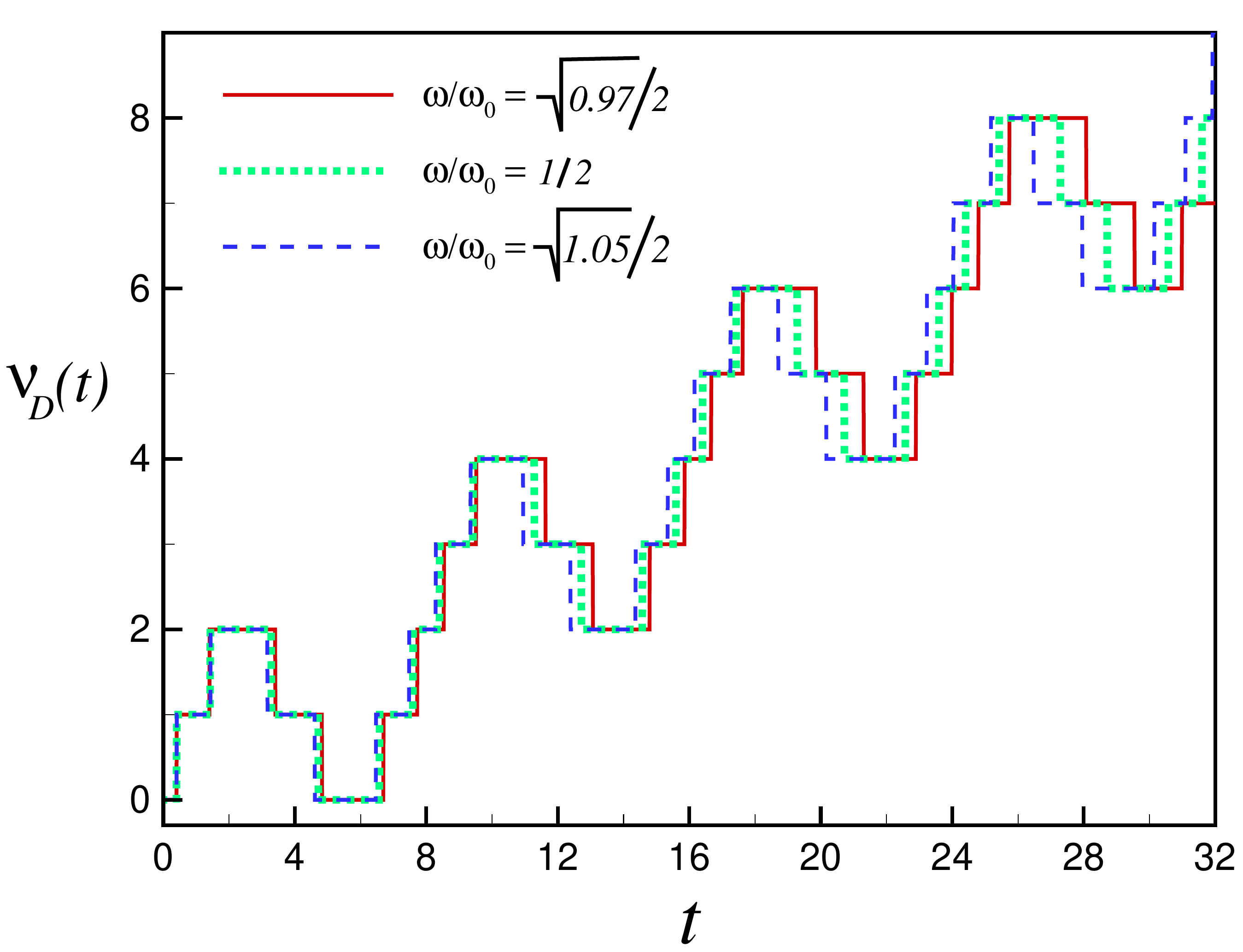}}
\centering
\caption{(Color online)
The dynamical topological order parameter
$\nu_{D}(t)$ of modulated Floquet XY model versus time for
$\gamma=1$, $h=\pi$, $J=h_{1}/2=\pi/6$, $\omega_{0}=\pi/2$ and
different ratio of driving frequencies close to commensurate case, namely:
$\omega/\omega_{0}=\sqrt{0.97}/2$, $\omega/\omega_{0}=1/2$, $(T_F=8)$,
and $\omega/\omega_{0}=\sqrt{1.05}/2.$}
\label{fig6}
\end{figure}
%
As mentioned before, the Hamiltonian of system (Eq. (\ref{eq1}))
is not periodic in time for incommensurate frequencies, which results in non-periodic dynamical topological quantum phase transition.
In Fig. \ref{fig6} the dynamical topological order parameter
of the model has been plotted for parameters, which  slightly deviate from the commensurate case.
As seen, within the
first driving period, the behaviour of dynamical topological order parameter for incommensurate cases
$\omega/\omega_{0}=\sqrt{0.97}/2, \sqrt{1.05}/2$
 are roughly the same as
that of the commensurate case, $\omega/\omega_{0}=1/2$.
The difference between dynamical topological order parameter of the incommensurate and commensurate cases increases by enhancing the driving period, which manifests the non-periodic behaviour of dynamical
topological quantum phase transition in the incommensurate case.
We observed in the incommensurate cases the return probabilities are non-decaying with time which should make them easier to trace in the laboratory though the dynamical
free energy displays non-periodic non-analyticities.

\bibliography{RefCFXYM}

\end{document}